\documentclass[conference]{IEEEtran}
\IEEEoverridecommandlockouts

\usepackage{cite}
\usepackage{amsmath,amssymb,amsfonts}
\usepackage{algorithm2e}
\usepackage{graphicx}
\usepackage{textcomp}
\usepackage{xcolor}
\usepackage{enumitem}
\usepackage{tabularx}
\usepackage{booktabs}
\usepackage{multirow}
\usepackage{svg}
\usepackage{multicol}
\usepackage{adjustbox}
\usepackage{comment}
\usepackage{subcaption}
\usepackage{url}
\usepackage{hyperref}
\usepackage[font=small,labelfont=bf]{caption}
\usepackage{cite}

\usepackage[numbers, sort&compress]{natbib}
\def\BibTeX{{\rm B\kern-.05em{\sc i\kern-.025em b}\kern-.08em
    T\kern-.1667em\lower.7ex\hbox{E}\kern-.125emX}}
    
\begin{document}

    \title{CVA6-CFI: A First Glance at RISC-V Control-Flow Integrity Extensions%
\thanks{This work was supported by the ISOLDE Chips JU
project (grant no. 101112274) and by Technology Innovation Institute, Secure Systems Research Center, Abu Dhabi, UAE (PO Box 9639).
Emanuele Parisi is supported by AI4S fellowship within the “Generación D” initiative, Red.es, Ministerio para la
Transformación Digital y de la Función Pública, for talent attraction (C005/24-ED CV1).
%Funded by the European Union NextGenerationEU funds, through PRTR.
%This work is part of the project PID2023-146511NB-I00 funded by the Spanish Ministry of Science, Innovation and Universities MCIU /AEI /10.13039/501100011033 and EU ERDF.
}}

\author{
    \IEEEauthorblockN{
        Simone Manoni\textsuperscript{*}, 
        Emanuele Parisi\textsuperscript{\dag}, 
        Riccardo Tedeschi\textsuperscript{*}, 
        Davide Rossi\textsuperscript{* \S}, 
        Andrea Acquaviva\textsuperscript{*}, 
        Andrea Bartolini\textsuperscript{*}
    }
    \IEEEauthorblockA{
        \textsuperscript{*}\textit{Department of Electrical, Electronic, and Information Engineering - University of Bologna, Italy} \\
        \textsuperscript{\dag}\textit{High Performance Domain-Specific Architectures Group - Barcelona Supercomputing Center, Spain} \\
        \textsuperscript{\S}\textit{Department of Digital Design and Open Hardware - Chips-IT, Italy} \\
        \{s.manoni, riccardo.tedeschi6, davide.rossi, andrea.acquaviva, a.bartolini\}@unibo.it, emanuele.parisi@bsc.es
        %Corresponding author: \texttt{\href{mailto:s.manoni@unibo.it}{s.manoni@unibo.it}}
    }
}

\maketitle
\begin{abstract}
This work presents the first design, integration, and evaluation of the standard RISC-V extensions for Control-Flow Integrity (CFI).
The \texttt{Zicfiss} and \texttt{Zicfilp} extensions aim at protecting the execution of a vulnerable program from control-flow hijacking attacks through the implementation of security mechanisms based on shadow stack and landing pad primitives.
We introduce two independent and configurable hardware units implementing forward-edge and backward-edge control-flow protection, fully integrated into the open-source CVA6 core. 
Our design incurs in only 1.0\% area overhead when synthesized in 22 nm FDX technology, and up to 15.6\% performance overhead based on evaluation with the MiBench automotive benchmark subset.
We release the complete implementation as open source.
\end{abstract}


    \section{Introduction}
\label{sec:Introduction}

RISC-V embedded platforms are experiencing rapid adoption across security and safety-critical industrial domains. 
These computing systems frequently execute critical workloads implemented in memory-unsafe languages, while operating in untrusted environments exposed over the network.
These features make them prime targets for cyberattacks~\cite{palka2024trends}. 
In fact, an attacker could exploit memory corruption bugs to modify vulnerable code pointers in memory, hijack program control-flow, and trigger arbitrary code execution by leveraging code-reuse attacks such as return-oriented programming (ROP), threatening the security of the whole system at risk~\cite{li2024sok, yang2021sok, shacham2007geometry}.
Control-Flow Integrity (CFI) has emerged as an effective defense mechanism against control-flow hijacking attacks by ensuring that programs execute only along intended control-flow paths, thereby preventing unauthorized deviations at runtime.
A typical CFI policy enforces protection on both “forward edges” and “backward edges” of control flow. 
Forward-edge policies protect indirect jumps where target addresses are computed dynamically during execution, such as those resulting from function pointer dereferencing in C. 
Conversely, backward-edge protection secures function returns, whose target addresses depend on values retrieved from the stack memory.
While various hardware-level approaches protect against control-flow diversion~\cite{kumar2020cats, dexie, sullivan2016efficient, titancfi}, dedicated ISA extensions have been the most adopted solution in commercial architectures like Intel and ARM, both featuring instructions devoted to the enforcement of CFI~\cite{liljestrand2019pacitup, shanbhogue2019security}.
RISC-V has recently ratified two CFI ISA extensions, \texttt{Zicfiss} and \texttt{Zicfilp}, which standardize an instruction-level mechanism for backward and forward-edge protection, implementing shadow-stack and landing pad CFI protection mechanisms, respectively.
To date, no study has evaluated their microarchitectural impact.
This paper explores how the RISC-V CFI extensions impact the performance and area of RISC-V application-class embedded cores.
\begin{itemize}
    \item We design two hardware components implementing the RISC-V CFI extensions: a shadow stack unit implementing the \texttt{Zicfiss} extension and a landing-pad unit implementing the \texttt{Zicfilp} extension.
    \item We integrate the designed components into CVA6 \cite{zaruba2019cost}, an embedded application-class core, presenting CVA6-CFI, a CVA6 variant with full CFI support. The complete implementation is available open-source\footnote{\url{https://github.com/AlSaqr-platform/cva6/tree/pulp-v1-culsans}}.
    \item We evaluate the impact of our RISC-V CFI extensions implementation, showing only 1.0\% area overhead and up to 15.6\% performance overhead across the MiBench automotive benchmark subset. Additionally, our implementation introduces the lowest area overhead compared to other hardware-centric CFI methods.
\end{itemize}

    \section{Background and Related Works}
\label{sec:Background and Related Works}

\subsection{Commercial ISA Extensions for CFI}

Both Intel and ARM have extended their ISAs with CFI instructions.
Intel's Control-flow Enforcement Technology (CET) includes two key features: Shadow Stack and Indirect Branch Tracking (IBT).
The Shadow Stack is a write-protected memory page dedicated to storing return address pointers whenever the core executes a \texttt{call} instruction.
The \texttt{ret} instructions raise an exception if the target address does not match the address stored in the shadow stack.
IBT introduced the \texttt{endbranch} instruction, which marks valid code targets for indirect calls and jumps. 
If an indirect control transfer attempts to target any invalid target, the processor raises an exception~\cite{shanbhogue2019security}.
ARM introduced the Branch Target Instructions (BTI) to protect indirect forward jumps and Pointer Authentication for function return protection.
BTI restricts indirect branches to target only a designated instruction, similar to Intel's \texttt{endbranch} working principle~\cite{yue2025efficient}.
Pointer authentication utilizes the unused upper bits of 64-bit pointers to store a Pointer Authentication Code (PAC).
During function calls, a PAC is generated from the return address and stored in the pointer's upper bits.
Before the function returns, the core authenticates the pointer, and an exception is triggered if an attacker tampers with the return address~\cite{liljestrand2019pacitup}.

\subsection{CFI RISC-V ISA extension}

Recently, RISC-V extended its ISA to support CFI through the introduction of Landing Pads and Shadow Stack extensions. 
The \texttt{Zicfilp} extension introduces a labeled landing pad instruction (\texttt{lpad}) encoded as \texttt{auipc x0, label}.
The execution of an indirect jump forces the processor into a state where the only instruction allowed to retire next is a \texttt{lpad} instruction with a matching label, or a software-check exception is triggered.
The expected label is pre-loaded in register \texttt{x7} and must match the label encoded in the immediate field of the \texttt{lpad} instruction.
The \texttt{Zicfiss} extension provides backward-edge protection via a shadow stack managed through dedicated instructions, including shadow stack push (\texttt{sspush}), pop and check (\texttt{sspchk}), pointer read (\texttt{ssrdp}), and atomic swap (\texttt{ssamoswap}).
Non-leaf functions save the link register into the shadow stack entry tracked by a dedicated shadow-stack pointer.
Programs maintain both the regular and shadow stacks to protect return addresses from tampering.
Unlike Intel's CET, the return address is not transparently pushed and popped from the shadow stack in conjunction with function calls and returns.

\subsection{CVA6 microarchitecture}

CVA6 is a 64-bit core designed for application-class systems.
It features an in-order, single-issue, six-stage pipeline with hardware support for virtualization.
The pipeline features two front-end stages responsible for generating the next program counter and interfacing with the instruction cache to fetch instructions. 
The decode stage realigns and decodes the instructions, storing them in the issue queue. 
The issue stage contains the issue queue, scoreboard, and reorder buffer, and it issues instructions to the execute stage once all operands are ready. 
The execute stage integrates the core functional units, and the commit stage retires up to two instructions per cycle, updates the register file, and resolves write-back conflicts through the reorder buffer.

    \section{Methodology}
\label{set:Methodology}

\subsection{Shadow Stack ISA extension design}

\begin{figure*}[ht]
    \centering
    \includegraphics[width=0.95\textwidth]{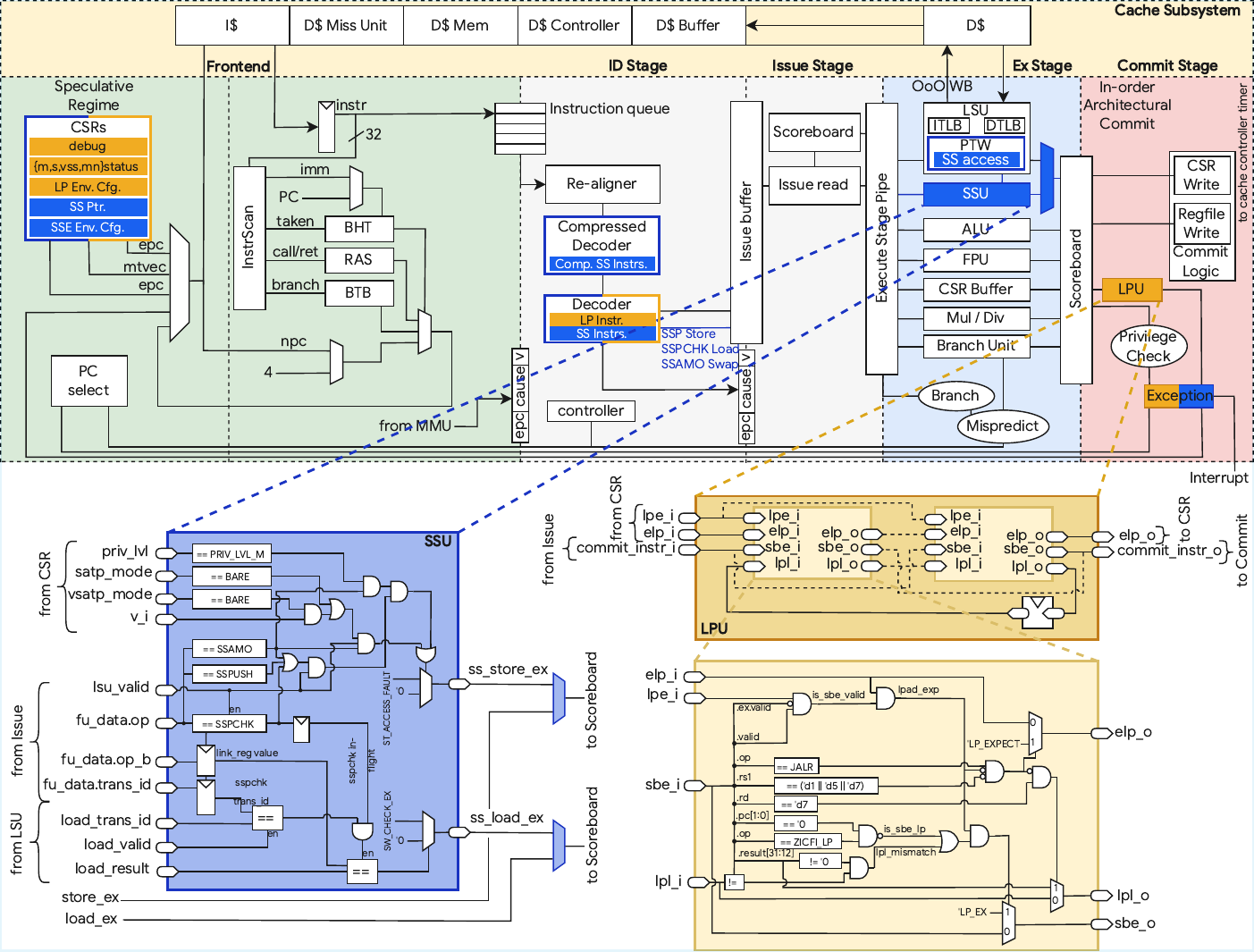}
    \caption{CVA6-CFI microarchitecture.}
    \label{fig:uarch}
    \vspace{-1em}
\end{figure*}

Figure \ref{fig:uarch} illustrates the CVA6 pipeline, enhanced with the CFI extensions.
The modules highlighted in blue have been introduced, or modified, to support the \texttt{Zicfiss} extension.
The Control Status Registers (CSRs) have been extended to include the Shadow Stack pointer, and the environment configuration registers for machine, supervisor, and hypervisor modes have been integrated and expanded with a Shadow Stack enable field. 
This field is utilized in the frontend to compute the Shadow Stack-enabled state across various privilege levels and is then forwarded to the decode and execute stages.
The decoder and compressed decoder are extended to handle the decoding of \texttt{sspush}, \texttt{sspchk}, \texttt{ssprd}, and \texttt{ssamoswap} instructions, as well as their compressed counterparts.
The \texttt{sspush} and \texttt{sspchk} instructions have been mapped to store and load instructions, respectively, with special operation tags to ensure that \texttt{Zicfiss} operations are differentiated from standard memory operations in later pipeline stages.
The same applies to the \texttt{ssamoswap} instructions, which are tagged with dedicated Shadow Stack swap W/D identifiers.
We have implemented a dedicated Shadow Stack Unit (SSU) in the execution stage, responsible for performing shadow stack pop checks and early instruction filtering before allowing the instructions to access the load-store unit (LSU).
If the SSU detects a \texttt{ssamoswap} operation executing in machine mode, or any \texttt{Zicfiss} instruction running with Supervisor or Virtual address translation and protection disabled, and at a privilege level below machine mode, a store access fault exception is raised and forwarded directly to the scoreboard. 
If neither of these two conditions is met, access to the LSU is allowed.
Furthermore, when the SSU detects a \texttt{sspchk} instruction, it buffers the link register content and the load transaction ID in dedicated registers and then waits for the load result. 
It also sets a flag to indicate that a \texttt{sspchk} is in progress.
When a \texttt{sspchk} is in-flight, the SSU compares the load transaction ID of each load result coming from the LSU with the buffered transaction ID.
If these IDs match, the SSU then checks the recorded link register value against the load result.
If there is a mismatch between these values, a software check exception is raised. 
Additionally, we extended the memory management unit (MMU) to enforce checks based on the type of page accessed.
\texttt{Zicfiss} instructions can access only shadow stack pages, while non-\texttt{Zicfiss} can access only non-shadow stack pages.
Any violation of this access policy results in a store access fault.

\subsection{Landing Pads ISA extension design}

The modules highlighted in orange in Figure \ref{fig:uarch} have been introduced, or modified, to support the \texttt{Zicfilp} extension.
To enforce forward-edge CFI, RISC-V compilers emit a dedicated \texttt{lpad} as the first instruction of any indirect jump targets. 
To support this mechanism, we integrate a Landing Pad Unit (LPU) at the interface between the scoreboard and the commit stage of the core pipeline. 
The LPU monitors ordered, non-speculative, and valid instructions, and it checks two events. 
First, when an update to the \texttt{x7} register is detected, the LPU records this as the most recent valid landing pad label configured by the program. 
Second, when an indirect jump is observed, the pipeline transitions to a state in which only the \texttt{lpad} instruction matching the current landing pad label is permitted to execute. 
Upon reaching this state, if a matching \texttt{lpad} instruction is encountered, it is retired without side effects; if not, the LPU raises an exception. 
By observing fully executed, non-speculative instructions, the LPU design facilitates seamless integration into pipelines such as CVA6, which feature multiple commit ports capable of retiring more than one instruction per cycle.
In these configurations, a chain of LPUs is instantiated, one per commit port, and each instance propagates information about \texttt{lpad} executions or landing pad label updates to the next unit.
While this design choice increases propagation delay in the commit stage, it addresses corner cases, such as when an indirect jump and its landing pad are retired in the same cycle.
Similarly to the \texttt{Zicfiss} extension, we extended the CSR register file and the decode stage with the necessary control and decoding logic to support the \texttt{Zicfilp} extension.

    \section{Results} 
\label{sec:Results}

\subsection{Hardware Characterization}
\begin{figure}
    \includegraphics[width=\columnwidth]{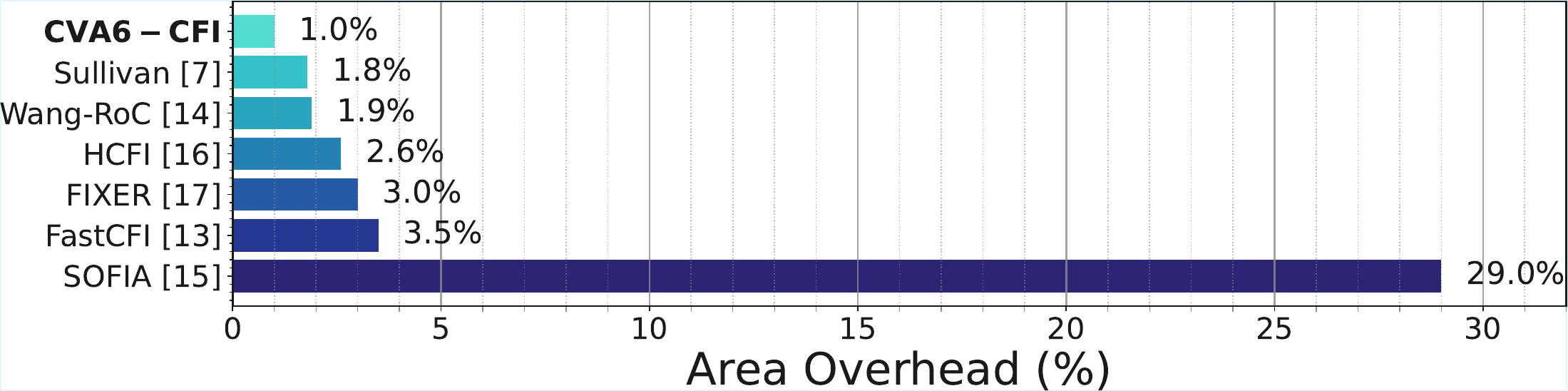}
    \caption{CVA6-CFI area overhead comparison.}
    \label{fig:areacomp}
    \vspace{-1em}
\end{figure}

To evaluate the impact on area and performance of the RISC-V ISA extensions, we synthesized the CVA6 core, with and without the \texttt{Zicfilp} and \texttt{Zicfiss} extensions, using Synopsys Design Compiler targeting GF22FDX technology.
The target operating frequency was fixed at 800~MHz under worst-case conditions (SSG corner, 0.72~V, –40°C to 125°C).
Table~\ref{tab:area} reports the impact of the CFI functional units on the core pipeline area.
The cache subsystem is excluded to avoid masking the overhead introduced by our CFI extensions, as the cache remains identical between the baseline CVA6 and the CFI-enhanced version.
Our implementation introduces a low area overhead across most pipeline stages, except for the commit stage.
Although this stage doubles in size due to the two LPU units, its relative contribution to the total pipeline area is low; therefore, the overall area increase for CVA6-CFI remains negligible, approximately equal to 1.0\%.
Moreover, the integration of the shadow stack and landing pad functionalities does not impact the core's operating frequency.
Figure~\ref{fig:areacomp}, extended from~\cite{hacfi}, shows the area overhead comparison between our proposed CFI design and prior hardware-centric CFI methods~\cite{feng2021fastcfi,hacfi,declercq2017sofia,sullivan2016efficient,christoulakis2016hcfi,de2019fixer}. Each overhead value is measured relative to its baseline architecture, with CVA6-CFI introducing the least overhead.
% \begin{table}[h!]
% \centering
% \caption{CVA6 Area evaluation}
% \label{area}
% \begin{tabular}{l c c c}
% \toprule
% \textbf{CVA6 Stage} & \textbf{Area Base (um)} & \textbf{Area w/ CFIext (um)} & \textbf{Area Overhead (\%)} \\ 
% \midrule
% CSR regfile & 7831 & 8220 & 4.97 \\
% Frontend \& ID & 1101 & 1124 & 2.06 \\
% Issue  & 25637 & 25788 & 0.59 \\
% Ex     & 61631 & 61854 & 0.36 \\
% Commit & 182 & 372 & 103.6 \\ \hline 
% Tot & 96382 & 97358 & 1.02 \\
% \bottomrule
% \end{tabular}
% \end{table}
\renewcommand{\arraystretch}{0.85}  % default is 1.0
\begin{table}[ht]
\centering
\caption{CVA6 Area Overhead Analysis in 22nm FDX}
\label{tab:area}
\begin{tabular}{lrrr}
\toprule
\multicolumn{1}{c}{\multirow{2}{*}{\vspace{-.2cm}CVA6 Pipeline}} &
\multicolumn{2}{c}{Area [$\mu m^2$]}                             &
\multicolumn{1}{c}{\multirow{2}{*}{\vspace{-.2cm}Overhead [\%]}} \\
\cmidrule(lr){2-3}
                                &
\multicolumn{1}{c}{Baseline}    &
\multicolumn{1}{c}{w/ CFI ext.} &
                                \\

\midrule
% CSR RegFile     &  7831 &  8220 &   4.97 \\
% Fetch \& Decode &  1101 &  1124 &   2.06 \\
% Issue           & 25637 & 25788 &   0.59 \\
% Execute         & 61631 & 61854 &   0.36 \\
% Commit          &   182 &   372 & 103.60 \\
CSR Reg. File   &  7831 &  8220 &   5.0 \\
Fetch \& Decode &  1101 &  1124 &   2.1 \\
Issue           & 25637 & 25788 &   0.6 \\
Execute         & 61631 & 61854 &   0.4 \\
Commit          &   182 &   372 & 103.6 \\
\cmidrule(lr){1-4}
% Total           & 96382 & 97358 &   1.02 \\
Total           & 96382 & 97358 &   1.0 \\
\bottomrule
\end{tabular}%
\end{table}

\subsection{Software Characterization}
\begin{figure}[t]
    \centering
    \begin{subfigure}{\columnwidth}
        \includegraphics[width=\textwidth]{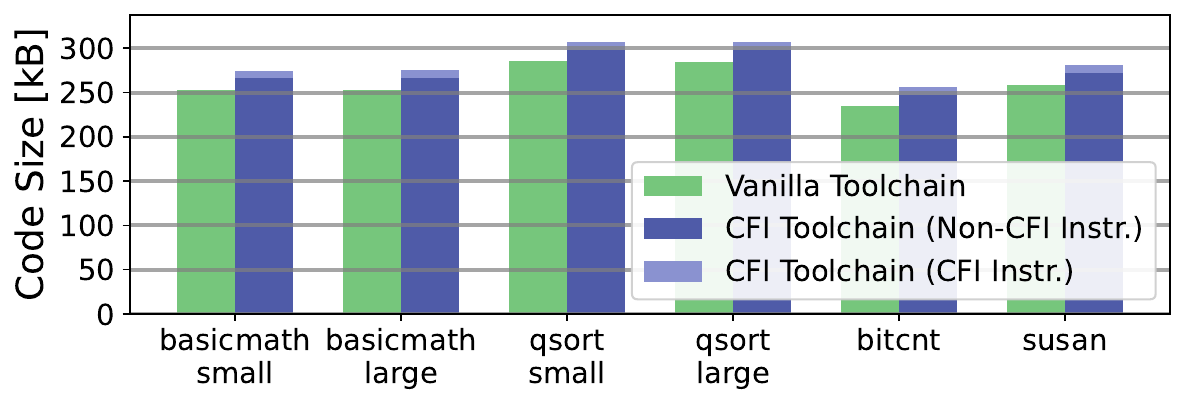}
        \vspace{-1.3em}
        \caption{Code size overhead for MiBench Automotive benchmarks.}
        \label{fig:codesize}
    \end{subfigure}

    \begin{subfigure}{\columnwidth}
        \includegraphics[width=\textwidth]{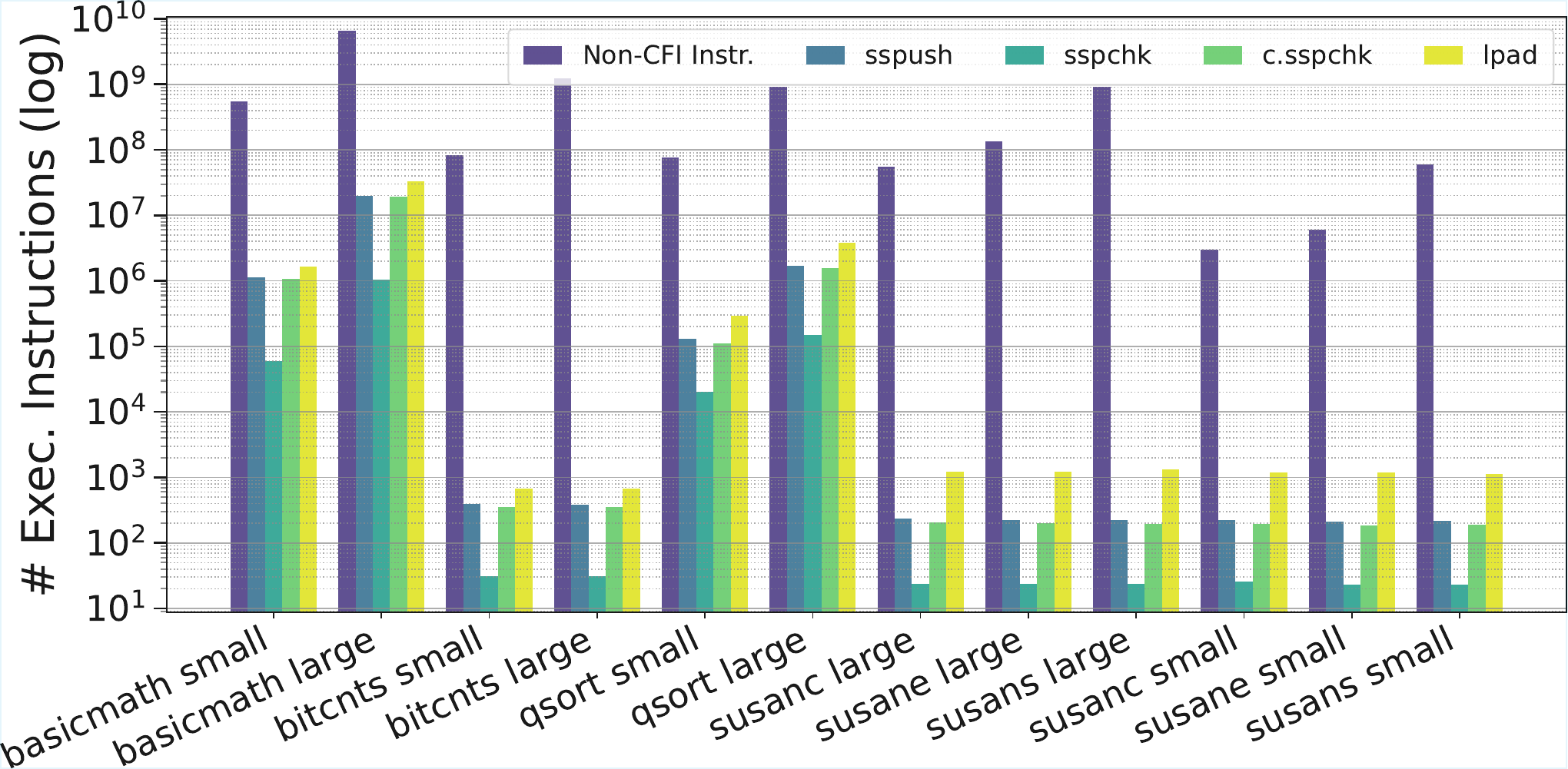}
        \vspace{-1.3em}
        \caption{Number of instructions executed for each MiBench program.}
        \label{fig:instrcount}
    \end{subfigure}
    
    \begin{subfigure}{\columnwidth}
        \includegraphics[width=\textwidth]{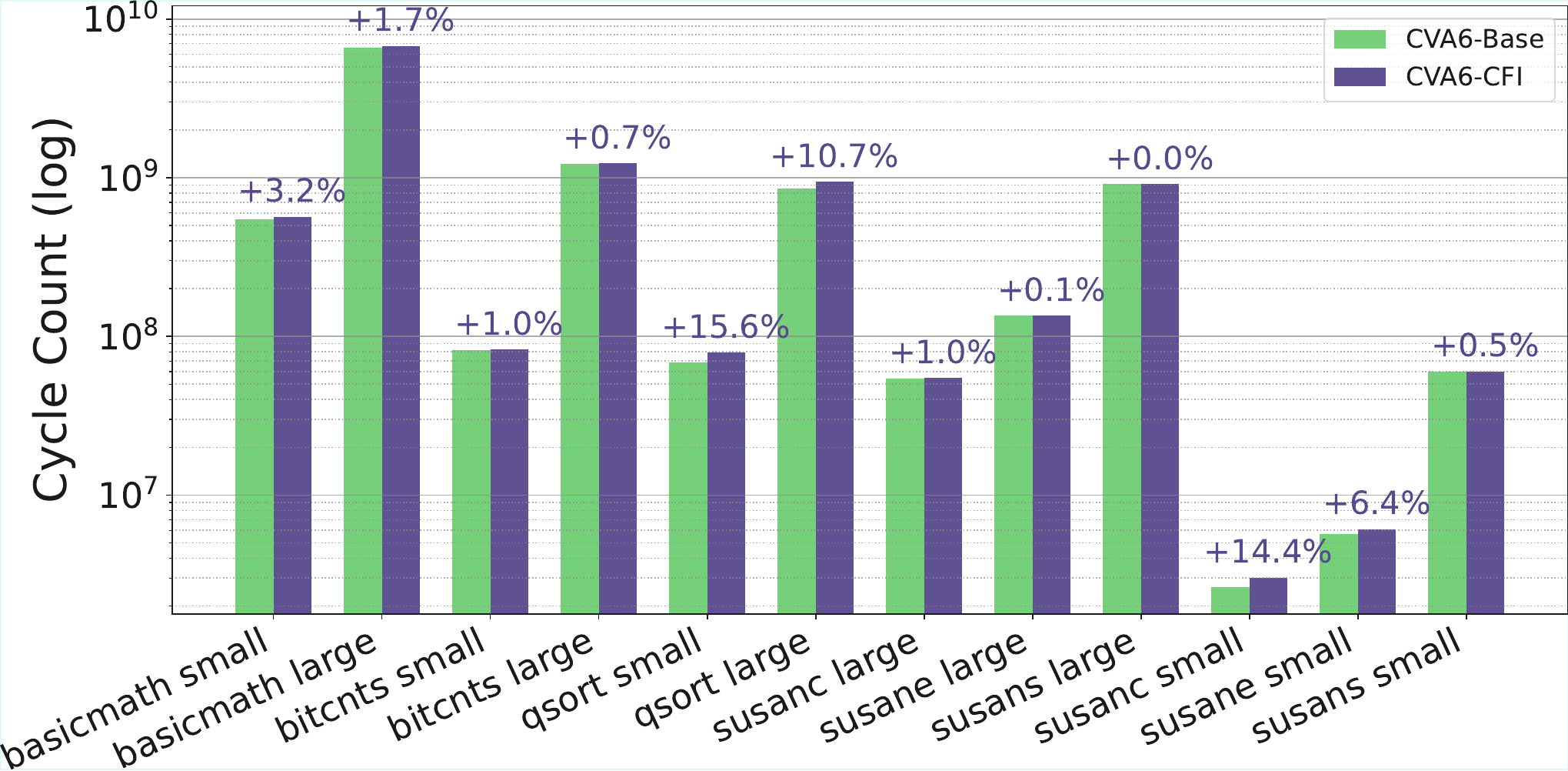}
        \vspace{-1.3em}
        \caption{Performance comparison between CVA6 baseline and CVA6-CFI.}
        \label{fig:latency}
    \end{subfigure}
    \caption{CVA6-CFI Software Characterization}
    \vspace{-1em}
\end{figure}

We evaluated the code size and runtime overhead of the RISC-V CFI extensions using the MiBench automotive benchmark subset~\cite{mibench} compiled with the CFI-aware SiFive GCC 13.3 toolchain\footnote{\url{https://github.com/sifive/riscv-gnu-toolchain/tree/cfi-dev}}.
Due to the absence of a stable CFI-patched Linux release, all benchmarks were executed on QEMU~\cite{qemu} in user-emulation mode, using a cycle-accurate model of the CPU calibrated against RTL simulations performed with Questa~2023.1.
Figure~\ref{fig:codesize} presents the code size overhead introduced by the CFI implementation across the MiBench automotive subset, compiled using the CFI-enabled toolchain. 
The results demonstrate consistent instrumentation costs with several key observations:
The CFI extensions introduce a stable absolute overhead of approximately 22-23~kB across all benchmarks, representing a relative increase of 7.6-9.2\%.
This consistency suggests that the instrumentation adds a largely fixed cost independent of specific application characteristics.
Notably, the overhead remains identical between small and large input variants of the same benchmarks (\textit{basicmath} and \textit{qsort}), indicating that the CFI cost is primarily determined by program structure rather than data size or complexity.
Figure~\ref{fig:instrcount} shows the instruction count for the selected MiBench programs, distinguishing between CFI and non-CFI instructions.
\textit{basicmath\_large} and \textit{qsort\_large} exhibit the highest CFI instruction counts, while \textit{bitcount} and \textit{susan} show minimal CFI activity.
This pattern directly correlates with function call density: \textit{basicmath\_large}'s deeply nested loops calling a cubic equation solver repeatedly generate intensive shadow stack activity.
In contrast, \textit{qsort\_large}'s recursive partitioning and frequent comparisons create substantial CFI overhead.
The \textit{bitcount}'s bitwise operations in tight loops and \textit{susan}'s image processing with direct function calls result in minimal CFI instructions execution.
Across all benchmarks, CFI instructions represent less than 0.5\% of total executed instructions.
Notably, \texttt{ssamoswap} instructions are absent from all benchmarks.
We assume they are primarily required in the Linux kernel for context switching and shadow stack management, so they are irrelevant to characterizing applications' behavior in the user space.
Similarly, \texttt{ssrdp} only appears once during the initial benchmark setup to read the shadow stack pointer for stack frame establishment. 
Therefore, we did not include them in the figure.
Figure~\ref{fig:latency} compares the cycle counts of programs compiled with the vanilla and CFI-enabled toolchains running on CVA6 and CVA6-CFI, respectively.
The CFI extension introduces a modest slowdown across all benchmarks, with a maximum relative overhead of 15.6\%.
This overhead is primarily determined by shadow stack operations (\texttt{sspush}, \texttt{sspchk}), which account for all CFI-related memory accesses.
Despite their prevalence, \texttt{lpad} instructions have a lower performance impact, as the LPU resolves them in a single cycle without requiring memory access in the execute stage. %when they reach the execute stage.
\begin{comment}
\begin{figure}[h]
  \includegraphics[width=\columnwidth]{figures/histogram_cfi_instr_only.pdf}
  \caption{Number of CFI instructions executed for each MiBench program.}
  \label{fig:instrcount}
\end{figure}
\begin{figure}[h]
  \includegraphics[width=\columnwidth]{figures/histogram_cycles_only.pdf}
  \caption{Latency comparison between CVA6 base and CVA6-CFI.}
  \label{fig:latency}
\end{figure}
\end{comment}
    \vspace{-0.5mm}
\section{Conclusion}
\vspace{-0.5mm}
This work presents CVA6-CFI, the first design and evaluation of the RISC-V standard extensions for CFI. 
Our approach introduces two independent hardware units implementing backward-edge and forward-edge control-flow protection. 
CVA6-CFI incurs in only 1.0\% of area overhead and up to 15.6\% performance overhead on the MiBench automotive subset; the full implementation is released open-source. 
Future work will aim at the silicon prototyping, power measurements and studying the impact of the RISC-V CFI extensions on advanced out-of-order and multi-thread core architectures.
\label{sec:Conclusion}

    \bibliographystyle{ieeetr}
    \bibliography{main}

\end{document}